\documentstyle[preprint,aps]{revtex}
\begin{document}
\title{Universal trapping scaling on the unstable manifold \\ for a
collisionless electrostatic mode}
\author{John David Crawford}
\address{Department of Physics and Astronomy\\
University of Pittsburgh\\
Pittsburgh, Pennsylvania  15260}
\date{\today}
\maketitle
\begin{abstract}
An amplitude equation for an unstable mode in a
collisionless plasma is derived from
the dynamics on the two-dimensional unstable manifold of the equilibrium. The
mode amplitude $\rho(t)$ decouples from the phase due to the
spatial homogeneity of the equilibrium, and the resulting one-dimensional
dynamics is analyzed using an expansion in $\rho$. As the linear growth rate
$\gamma$ vanishes, the expansion coefficients
diverge; a rescaling $\rho(t)\equiv\gamma^2\,r(\gamma t)$ of the mode
amplitude absorbs these singularities and reveals that the mode electric field
exhibits trapping scaling $|E_1|\sim\gamma^2$ as $\gamma\rightarrow0$.
The dynamics for $r(\tau)$ depends
only on the phase $e^{i\xi}$ where
$d\epsilon_{{k}} /dz=|{\epsilon_{{k}}}|e^{-i\xi/2}$ is the
derivative of the dielectric as $\gamma\rightarrow0$.
\end{abstract}


The collisionless evolution of an unstable electrostatic mode is a fundamental
topic in the theory of strong wave-particle interactions. The linear
instability arises from a resonant interaction between an initial electrostatic
fluctuation and particles at the phase velocity of the linear mode; the
nonlinear evolution is marked by the trapping of the resonant particles in the
wave potential and
decaying oscillations in the wave amplitude due to the bouncing and phase
mixing
of the trapped particles. The difficulty of treating the dynamics of this
process analytically is well known, and the extensive literature tends to
focus on certain special regimes where simplifying approximations are
possible; for example, instabilities due to a cold low density beam\cite{dru}
or  a gentle ``bump on tail'' \cite{fried}-\cite{sim2}.

Early work on the interaction of a narrow spectrum of weakly unstable waves
with a cold electron beam identified the importance of particle trapping and
predicted a relation $\omega_b\sim\gamma$ between the linear growth rate
$\gamma$ and the bounce frequency $\omega_b^2=ekE_k/m$ in the nonlinear
state that emerges after the linear instability has saturated.\cite{dru} This
relation is more transparently stated as a property of the electric field $E_k$
of the saturated wave: $E_k\sim\gamma^2$ as $\gamma\rightarrow0$; a property we
refer to as trapping scaling for $E_k$. Initial studies of the saturated state
for the bump on tail instability also predicted trapping
scaling.\cite{fried,dewar} These investigations all invoke approximations
treating the response of the non-resonant electrons as linear or adiabatic.

By contrast, Simon and Rosenbluth constructed a time-asymptotic state by
perturbatively expanding the Vlasov equation in the beam density and demanding
that secular terms vanish at each order. Their procedure led to expressions
involving singular functions which were defined by prescribing certain
regularization procedures; the resulting theory predicted nonlinear states with
much larger electric fields $E_k\sim\sqrt{\gamma}$.\cite{sim1} Subsequent
perturbative calculations by other groups involve similar assumptions and reach
the same basic  conclusion\cite{janssen,burnap} with one exception: Larsen has
studied the bump on tail problem using a multiple scale expansion in
time and velocity; in his formulation, a ``singular layer'' at the phase
velocity leads him to posit trapping scaling as an ansatz.\cite{larson}
Numerical simulations of the instability find the initial, possibly metastable,
nonlinear state with trapping scaling\cite{den,sim2} but have sometimes claimed
to detect a slow growth in $E_k$ on very long time scales.\cite{sim2} Recent
laboratory
experiments of an electron beam interacting with an electrostatic wave
(supported by a travelling wavetube) find the nonlinearly saturated wave
amplitude is described by trapping scaling over the length scale of the
experiment.\cite{tsu}

In this paper we describe a new approach which also treats the Vlasov equation
perturbatively, but simplifies the problem in a new way: the initial conditions
are restricted so that the evolution occurs on the unstable manifold of the
equilibrium $F_0$.\cite{rigor} Physically this restriction means that only the
unstable mode is initially excited, rather than viewing the mode as one
component of an arbitrary fluctuation. Mathematically the unstable manifold
provides a finite-dimensional setting which partially compensates for
the absence of a finite-dimensional center manifold in this problem. In the
simplest case, the unstable manifold defines a two-dimensional problem where
the invariance of $F_0$ under spatial translation implies that the evolution of
the mode amplitude decouples from the phase and is described by a
one-dimensional dynamical system. We formulate and analyze this one-dimensional
problem.

The dimensionless equations for the potential $\phi(x,t)$ and electron
distribution function $F(x,v,t)$ are
\begin{equation}
\partial_t F+v\partial_x F+\partial_x\phi\,\partial_v F=0\hspace{0.25in}
\partial_x^2 \phi=\int^\infty_{-\infty}dv\,F\;-1\label{vlasov}
\end{equation}
where
\begin{equation}
\int^{L/2}_{-L/2}dx\,\int^\infty_{-\infty}dv\,F(x,v,t)=1.\label{norm}
\end{equation}
Given an equilibrium $F_0(v,\mu)$, Eq.(\ref{vlasov}) gives the
evolution equation for $f(x,v,t)\equiv F(x,v,t)-F_0(v,\mu)$,
\begin{equation}
\partial_t f={\cal{L}} f+{\cal{N}}(f)\label{pertvlasov}
\end{equation}
where ${\cal{N}}(f)=-\partial_x\phi\,\partial_v f$ and ${\cal{L}}
f=-v\partial_x
f-\partial_x\phi\,\partial_v F_0$; here $\mu$ denotes
parameters such as density or temperature that determine the properties of
$F_0$. We assume periodic boundary conditions on $f$ thus the Fourier
components $f_k(v)\,e^{ikk_cx}$ are discrete multiples of $k_c=2\pi/L.$

The length of the system is chosen so that as $\mu$ varies an instability
occurs for $k_c$ (or $k=1$) corresponding to an eigenfunction
${\cal{L}}\Psi=\lambda\Psi$ where $\Psi(x,v)=e^{ik_cx}\,\psi_c(v)$,
$\lambda=-ik_cz_0$, and $\psi_c(v)=\partial_vF_0/k_c^2(v-z_0)$. The unstable
mode is determined by a root $z_0=v_p + {i\gamma}/{k_c}$ of the dielectric
function
\begin{equation}
\epsilon_{{k}}(z)\equiv 1- \frac{1}{(kk_c)^2}\int^\infty_{-\infty}\,dv\,
\frac{\partial_vF_0(v,\mu)}{v-z}\hspace{0.5in}(\mbox{\rm Im}\;z>0)
\label{diefcn}
\end{equation}
and both $v_p$, the phase velocity of the mode, and $\gamma$ depend on the
equilibrium parameters $\mu$. For our purposes it is not important exactly what
parameters $\mu$ represents or how they are varied, the mode is assumed to be
neutrally stable for $\mu=\mu_c$ and to become unstable if $\mu$ is
appropriately shifted away from $\mu_c$; the inverse limit
$\mu\rightarrow\mu_c$ from the unstable regime will be denoted by
$\gamma\rightarrow0^+$. In addition we assume the derivatives
$d^n\epsilon_{{k}}
(z_0)/dz^n$ have finite limits as $\gamma\rightarrow0^+$ and that the first
derivative ${\epsilon_{{k}} '}(z_0)$ is non-zero. This latter requirement that
$z_0$ is
a simple root will typically be satisfied when a single parameter is varied,
e.g. beam density or beam velocity.

When $F_0(v,\mu)$ lacks reflection symmetry as in a beam-plasma system then
$v_p\neq0$ and the eigenvalue $\lambda$ is complex. For $\gamma>0$ there is a
complex conjugate pair of eigenvalues $(\lambda,\lambda^\ast)$ in the right
half plane and a symmetrically placed pair $(-\lambda,-\lambda^\ast)$ in the
left half plane; as $\gamma\rightarrow0^+$ this eigenvalue quadruplet merges
into the continuous spectrum on the imaginary axis.\cite{crawhis} For
reflection-symmetric problems, $F_0(v,\mu)=F_0(-v,\mu)$, the eigenvalue can be
real or complex; for example, two-stream instabilities correspond to real
eigenvalues of multiplicity two: $\Psi$ and $\Psi^\ast$ are linearly
independent eigenvectors for $\lambda$.\cite{cowley} An instability with
complex $\lambda$ leads to four-dimensional unstable manifolds when $F_0$ is
reflection-symmetric and will not be considered here.\cite{jdcaps}

For $\lambda$ complex (no reflection symmetry) or $\lambda$ real (with
reflection symmetry), we decompose $f$ to isolate the two unstable modes
\begin{equation}
f(x,v,t)=\left[A(t)\Psi(x,v) + cc\right] + S(x,v,t)\label{linmodes}
\end{equation}
where $A(t)=(\tilde{\Psi},f)$ and $(\tilde{\Psi},S)=0$; $\tilde{\Psi}$ is the
adjoint eigenvector corresponding to $\lambda^\ast$ and $(G_1,G_2)\equiv
\int\,dx\int\,dv\, G_1(x,v)^\ast G_2(x,v)$ denotes the inner product. The
equations for $A$ and $S$ follow from (\ref{pertvlasov})
\begin{eqnarray}
\dot{A}&=&\lambda\,A+(\tilde{\Psi},{\cal{N}}(f))\label{Adot}\\
\partial_t S&=&{\cal{L}} S+{\cal{N}}(f)-\left[(\tilde{\Psi},{\cal{N}}(f))\,\Psi
+ cc\right];
\label{Sdot}
\end{eqnarray}
the  linear terms are now decoupled, but nonlinear couplings between $\dot{A}$
and $\partial_t S$ remain. For $\gamma>0$, the modes $\Psi$ and $\Psi^\ast$
span the two-dimensional unstable subspace $E^u$ which is invariant under the
linear flow $\partial_t f={\cal{L}} f$. The nonlinear terms couple the unstable
modes to the two stable modes (spanning the stable subspace $E^s$) and to the
continuum (which spans the infinite-dimensional center subspace $E^c$); these
interactions bend the unstable subspace into a two-dimensional unstable
manifold which is invariant for the full nonlinear evolution. Solutions on this
manifold asymptotically approach $F_0$ at an exponential rate $e^{\gamma t}$ as
$t\rightarrow-\infty$.

At the equilibrium, the manifold is tangent to $E^u$ and thus can be
represented as the graph of a function $H: E^u\rightarrow E^c\oplus E^s$; see
Figure 1. With respect to the decomposition in (\ref{linmodes}), when $f$ is a
point on the unstable manifold (denoted $f^u$) then
\begin{equation}
f^u(x,v)=\left[A\Psi(x,v) + cc\right] + H(x,v,A,A^\ast);\label{unstablef}
\end{equation}
hence the time dependence of $S$ for a solution on the manifold is determined
by the dynamics of $A(t)$ and the geometry of the manifold:
$S(x,v,t)=H(x,v,A(t),A^\ast(t))$. For such a solution $\partial_t
S=\dot{A}\,\partial_{A}H+\dot{A}^\ast\,\partial_{A^\ast}H$ and consistency with
(\ref{Sdot}) requires
\begin{equation}
\left[\dot{A}\,\partial_{A}H+\dot{A}^\ast\,\partial_{A^\ast}H\right]
{\rule[-3.0mm]{0.25mm}{8.0mm}}_{f=f^u}
={\cal{L}} H+{\cal{N}}(f^u)-\left[(\tilde{\Psi},{\cal{N}}(f^u))\,\Psi +
cc\right].\label{Heqn}
\end{equation}
Solving this equation for $H$, then determines the representation of $f^u$ in
(\ref{unstablef}). Setting $f=f^u$ in (\ref{Adot}) yields the desired
description of the dynamics on the unstable manifold:
\begin{equation}
\dot{A}=\lambda\,A+(\tilde{\Psi},{\cal{N}}(f^u))\equiv
V(A,A^\ast,\mu);\label{unAdot}
\end{equation}
this is an autonomous two-dimensional dynamical system for $A(t)$. For small
$A$, the graph function $H$ is second order in $A$ and $f^u\approx
[A(t)\Psi(x,v) + cc]$, in this sense the dynamics on the unstable manifold
corresponds to an initial excitation of only the unstable modes.

The spatial translation symmetry of the
problem constrains $V$ to have the form $V(A,A^\ast,\mu)=A\,p(|A|^2,\mu)$ where
$p$ is an undetermined function. If
$F_0$ is reflection-symmetric as in the two-stream instability, then $p$ must
be real-valued; otherwise $p$ is complex-valued. Finally in amplitude/phase
notation $A=\rho e^{-i\theta}$ (\ref{unAdot})
becomes
\begin{equation}
\dot{\rho}=\rho\,\mbox{\rm Re}\,[p(\rho^2,\mu)]\hspace{0.75in}
\rho\dot{\theta}=-\mbox{\rm Im}\,[p(\rho^2,\mu)];\label{amphase}
\end{equation}
thus on the unstable manifold the evolution of the mode amplitude $\rho(t)$
decouples from the phase $\theta(t)$ and is a one-dimensional problem.

For small wave amplitudes, we have investigated the properties of
(\ref{amphase}) by representing $p(\rho^2,\mu)$ as a power series,
\begin{equation}
p(\rho^2,\mu)=\sum^\infty_{j=0}\,p_j(\mu)\,\rho^{2j}.\label{series}
\end{equation}
Clearly $p_0(\mu)=\lambda$ from (\ref{Adot}) and the higher order coefficients
$p_j(\mu)$ are determined by solving (\ref{Heqn}) for $H$ as a power series in
$(A,A^\ast)$ then substituting into (\ref{unAdot}); this calculation will be
presented elsewhere.\cite{jdc} Of greatest interest
are the properties of $p_j$ in the regime of weak instability
$\gamma\rightarrow0^+$: for $j>0$, these coefficients are singular at every
order
\begin{equation}
p_j(\mu)=\frac{b_j(\mu_c)}{\gamma^{4j-1}}\,[1+{\cal{O}}(\gamma)],\label{sing}
\end{equation}
and the remaining asymptotic dependence on the equilibrium $F_0$ is remarkably
simple.
Each $b_j(\mu_c)$ depends on $F_0$ only through the phase
$e^{i\xi(\mu_c)}\equiv\,{\epsilon_{{k}} '}(v_p)^\ast/{\epsilon_{{k}} '}(v_p)$,
which is determined by
the limiting value of ${\epsilon_{{ k}}'}(z_0)$. More precisely, at each order
there is
a calculable function
$Q_j$, independent of $F_0$, such that
\begin{equation}
b_j(\mu_c)=Q_j(e^{i\xi(\mu_c)});\label{bj}
\end{equation}
in particular, we find $b_1(\mu_c)=-1/4$ although in general $Q_j$ will not be
constant  at higher order.

The significance of the
divergence in (\ref{sing}) is clearer when we introduce a rescaled mode
amplitude
$\rho(t)\equiv\gamma^2\,r(\gamma t)$ which varies on the slow time scale
$\tau\equiv\gamma t$ and rewrite the dynamics (\ref{amphase}) as
\begin{equation}
\frac{dr}{d\tau}=r\left\{1+\sum^\infty_{j=1}\,
\mbox{\rm
Re}\;\left[b_j(\mu_c)+{\cal{O}}(\gamma)\right]\,r^{2j}\right\}\hspace{0.55in}
\dot{\theta}=\omega-
\gamma\,\sum^\infty_{j=1}\,\mbox{\rm
Im}\;\left[b_j(\mu_c)+{\cal{O}}(\gamma)\right]\,r^{2j}
\label{rescaled}
\end{equation}
where $\omega=k_cv_p$ is the mode frequency. The $\gamma\rightarrow0^+$ limit
for (\ref{rescaled}) is nonsingular and the asymptotic equations have some
notable features. First, the dependence on $F_0$ is entirely contained in the
three parameters $\gamma, \omega,$ and $e^{i\xi}$; in particular at $\gamma=0$
the rescaled amplitude dynamics depends only on $e^{i\xi}$. If this phase is
fixed, then any variations in densities or temperatures characterizing $F_0$ do
not affect the evolution of
$r(\tau)$. For example, a beam-plasma instability (complex $\lambda$) and a
two-stream instability (real $\lambda$), compared at a fixed value of
$e^{i\xi}$, have identical amplitude equations up to ${\cal{O}}(\gamma)$
corrections.
A second feature is that the linear term defines a growth rate that is not
small, i.e. not ${\cal{O}}(\gamma)$, rather the growth rate is unity. Moreover,
unless
$b_j$ happens to vanish, {\em all} the higher order terms $r^{2j}$ in the
amplitude equation are order unity in the limit $\gamma\rightarrow0^+$. Thus
there is no small parameter in (\ref{rescaled}) to justify a truncation of the
series, and consequently it is not straightforward to calculate the
time-asymptotic amplitude $r(\tau\rightarrow\infty)$. However, assuming that
$r(\tau)$ approaches a limiting value $r_\infty$ as $\tau\rightarrow\infty$,
then it follows that this time-asymptotic state is a BGK mode.\cite{jdc}

The implications of (\ref{sing}) for the behavior of the electric field $|E_1|$
at $k_c$ (or $k=1$) follow from $ik_cE_1=\int dv\,f_1(v,t)=A(t)+\int
dv\,H_1(v,A,A^\ast)$; one can show that
\begin{equation}
k_c\,|E_1(t)|=
\gamma^2\,r\,[1+r^2\,\hat{\Gamma}_1(r^2,e^{i\xi})+{\cal{O}}(\gamma)]
\label{tscaling}
\end{equation}
where $\hat{\Gamma}_1$ represents a power series in $r^2$ whose coefficients
depend on $F_0$ only through $e^{i\xi}$.\cite{jdc} This expression predicts
that the trapping scaling $|E_1|\sim\gamma^2$ is a universal characteristic of
the entire evolution; in particular it should hold for the time-asymptotic
state. This conclusion agrees with the early work on the instability due to a
small cold beam,\cite{dru} and generalizes it to a much wider class of
instabilities.

An interesting aspect of (\ref{rescaled}) - (\ref{tscaling}) is the apparent
absence of the familiar trapping oscillations in the evolution of $|E_1|$. This
evolution is determined by $r(\tau)$ whose dynamics is described by an
autonomous one-dimensional flow, and it is well known that smooth
one-dimensional equations cannot describe oscillations. We have conjectured
elsewhere\cite{jdc1}, on the basis of simpler exactly solvable models, that the
unstable manifold develops a spiral structure away from the equilibrium as
illustrated in Figure 2. If this is correct, then representing the dynamics on
the manifold via a mapping $H$ from the unstable subspace yields a vector field
on $E^u$ with branch point singularities at the points where the flow moves
from one branch of the spiral to the next. Thus $\mbox{\rm
Re}\,[p(\rho^2,\mu)]$  would have a branch point $\mbox{\rm
Re}\,[p(\rho^2,\mu)]\sim\sqrt{\rho^2_b-\rho^2}$ at $\rho^2_b$ and as the mode
grows $\rho^2(\tau)\rightarrow\rho^2_b$ would signal the onset of trapping
oscillations with the passage of the trajectory to the next branch of the
unstable manifold. Note that a trajectory will reach such a node in finite
time, unlike the more familiar situation of a node where the vector field is
differentiable and the approach time is infinite. In addition the loss of
smoothness at $\rho=\rho_b$ introduces the lack of uniqueness needed by the
solution to pass through the branch point. Such a spiral structure
would present a significant obstacle to using the power series (\ref{series})
to determine the time-asymptotic amplitude $r_\infty$.

The coefficients $p_j(\mu)$ are calculated as integrals over velocity and the
singular behavior in (\ref{sing}) arises from  pinching singularities that
develop at $v=v_p$ as $\gamma\rightarrow0^+$. The regularization procedures
proposed in previous treatments would modify these integrals to remove the
pinching singularity and eliminate the divergences reflected in (\ref{sing}).
Any such regularization replaces trapping scaling by $|E_1|\sim\sqrt{\gamma}$
which is typical of a Hopf or pitchfork bifurcation in which there is no
continuous spectrum on the imaginary axis. In these latter bifurcations the
coefficients  $p_j$ are nonsingular, the series for $p$ can be truncated, and
time-asymptotic state found by balancing the linear term against a cubic
nonlinearity. However, our calculations show that the singularities in $p_j$
found here simply imply a different dependence of $\rho(t)$ and $|E_1|$ on
$\gamma$ as $\gamma\rightarrow0^+$, and that there is no need to introduce an
ad hoc regularization.

Instabilities in other Hamiltonian systems, including ideal shear
flows\cite{case} and solitary waves\cite{pegowein1}, also exhibit key features
of this problem, most notably that the unstable modes correspond to
eigenvalues emerging from a neutral continuum at onset. It will be interesting
to determine if similar singularities arise in the amplitude equations for the
unstable modes in these problems.\cite{jdcek} By contrast, there is at least
one example, a phase model for the onset of synchronized behavior in a
population of oscillators, in which the critical eigenvalues emerge from the
continuum at the onset of instability but the amplitude equations are
nonsingular and $\sqrt\gamma$ scaling is found (at least in the best understood
case of a real eigenvalue).\cite{sm,jdc2} This difference in the nonlinear
behavior seems noteworthy since the linear dynamics of the model is
qualitatively similar to Vlasov although apparently lacking a Hamiltonian
structure.\cite{smm}

\begin{figure}
\caption{Local geometry of the unstable manifold; the equilibrium $F_0$ is at
the origin}
\label{fig1}
\end{figure}
\begin{figure}
\caption{Conjectured spiral structure in the global unstable manifold}
\label{fig2}
\end{figure}


\begin{references}
\bibitem{rigor} Our appeal to invariant manifold theory is heuristic and serves
to motivate the perturbation theory. Mathematically, it seems to be an open
problem to rigorously construct invariant manifolds for Vlasov equilibria.
\end{references}
\end{document}